\documentclass[conference]{IEEEtran}
\IEEEoverridecommandlockouts
\usepackage{cite}
\usepackage{amsmath,amssymb,amsfonts}
\usepackage{algorithmic}
\usepackage{graphicx}
\usepackage{textcomp}
\usepackage{xcolor}
\usepackage{verbatimbox}
\newcommand{\etal}{\textit{et al}., }
\def\BibTeX{{\rm B\kern-.05em{\sc i\kern-.025em b}\kern-.08em
    T\kern-.1667em\lower.7ex\hbox{E}\kern-.125emX}}
\usepackage[bb=dsserif]{mathalpha}
\usepackage{xurl}

\begin{document}
\title{Resurrection Attack: Defeating Xilinx MPU's Memory Protection
\thanks{This work has been supported in part by a grant from the National Science Foundation (NSF).}
}

\author{\IEEEauthorblockN{Bharadwaj Madabhushi, Chandra Sekhar Mummidi, Sandip Kundu, Daniel Holcomb}
\IEEEauthorblockA{\textit{Department of Electrical and Computer Engineering} \\
\textit{University of Massachusetts Amherst}\\
\{\textit{bmadabhushi, cmummidi, kundu, dholcomb}\}@umass.edu}
}

\maketitle

\begin{abstract}
Memory protection units (MPUs) are hardware-assisted security features that are commonly used in embedded processors such as the ARM 940T, Infineon TC1775, and Xilinx Zynq. MPUs partition the memory statically, and set individual protection attributes for each partition.
MPUs typically define two protection domains: user mode and supervisor mode. Normally, this is sufficient for protecting the kernel and applications.
However, we have discovered a way to access a process memory due to a vulnerability in Xilinx MPU (XMPU) implementation that we call \textit{Resurrection Attack}. 
We find that XMPU security policy protects user memory from unauthorized access when the user is active. However, when a user's session is terminated, the contents of the memory region of the terminated process are not cleared. 
An attacker can exploit this vulnerability by gaining access to the memory region after it has been reassigned. The attacker can read the data from the previous user's memory region, thereby compromising the confidentiality.
To prevent the Resurrection Attack, the memory region of a terminated process must be cleared. However, this is not the case in the XMPU implementation, which allows our attack to succeed.
The Resurrection Attack is a serious security flaw that could be exploited to steal sensitive data or gain unauthorized access to a system. It is important for users of Xilinx FPGAs to be aware of this vulnerability until this flaw is addressed.

\end{abstract}


\section{Introduction}

Field-Programmable Gate Arrays (FPGAs) are versatile integrated circuits that can be programmed and reconfigured to perform any logic function, from simple logic operations to complex computations. FPGAs can be used to implement algorithms directly in hardware, which can often lead to greater performance and power efficiency. Consequently, FPGAs are increasingly being used in various applications, including networking, critical infrastructure, aerospace, defense, and finance. FPGAs are also being used in high-performance cloud computing systems such as Amazon EC2 F1 instances \cite{amazonAmazonInstances} due to their efficiency and programmability even during runtime. From a security perspective, the ability to dynamically reconfigure FPGAs, also known as runtime update support, can be a major security vulnerability.
 


There have been several studies on security of FPGAs \cite{naghibijouybari2022microarchitectural,zeitouni2020sok}. They primarily focus static attack vectors such as supply chain vulnerabilities \cite{rovzic2017monte}, malicious logic insertion \cite{cruz2023framework}, Trojans \cite{dhavlle2021design}, backdoor attacks on FPGAs including timing violation induced faults, replay attacks \cite{mahmoud2019timing,zhang2015reconfigurable}, and bit-stream tampering \cite{kim2019safedb}. 

There has also been studies on physical attacks on FPGAs. One type of physical attack on FPGAs is a remote power side-channel attack where an attacker can rent an FPGA instance, build power monitors, and use a power analysis attack to steal secret information \cite{zhao2018fpga}. Attackers can reverse engineer FPGA logic by analyzing the bitstream, a binary file containing the configuration data \cite{maes2011pay,zhang2019comprehensive,yu2018recent,narayanan2023reverse}. Fault-injection attacks aim to exploit vulnerabilities in the FPGA design by injecting faults into the system \cite{michel2017seu,benevenuti2021robust,shirazi2013fast}. 

Other attacks target FPGA interface components. For example, Weissman \etal performed a row-hammer attack on CPU main memory from the FPGA \cite{weissman2019jackhammer} and Ye \etal exploited the new attack surface between CPU-FPGA system \cite{ye2018hisa}. Tin \etal demonstrated how PCIe contention can be used to attack the security of FPGAs in cloud data centers. They showed that by identifying instances of PCIe contention among FPGA slots, they could accurately correlate co-located FPGAs and their corresponding instance allocations. This information could then be used to launch other attacks \cite{tian2021cloud}. Giechaskiel \etal showed how PCIe contention could be used to establish covert and side channels for covert transmission of information between virtual machines (VMs) \cite{giechaskiel2021cross}.

There has also been studies on software based attacks. For example, malwares running on CPU can access the Block RAM (BRAM) through Direct Memory Access (DMA), and a hardware Trojan in FPGA can leak or modify video output frames of the CPU memory \cite{ye2018hisa}.

\subsection{Multi-tenant FPGA vulnerabilities}
In Multi-tenant FPGAs, a single FPGA fabric is shared between multiple users by partial reconfiguration, thereby increasing the utilization of FPGA logic. However, the adoption of multi-tenancy introduces novel security susceptibilities. Despite  isolation mechanisms to keep user logic instances separate, attackers can exploit  shared electrical components and attack co-located applications. Giechaskiel \etal used the FPGA interconnect to leak information using the observation of long wires that carry logical 1 reduce propagation delay of unconnected wire \cite{giechaskiel2018leaky}. Similarly, Gnand et al. utilized shared power distribution networks among tenants to construct a covert communication channel bridging logically isolated users \cite{gnad2021voltage}.  These studies underscore how spatially shared FPGA fabrics can be rendered vulnerable to the creation of side or covert channels through the exploitation of shared resources. In temporal sharing, FPGA fabric is shared by users in different time slots.

\begin{figure}[t!]
    \centering
    \includegraphics[width=\linewidth]{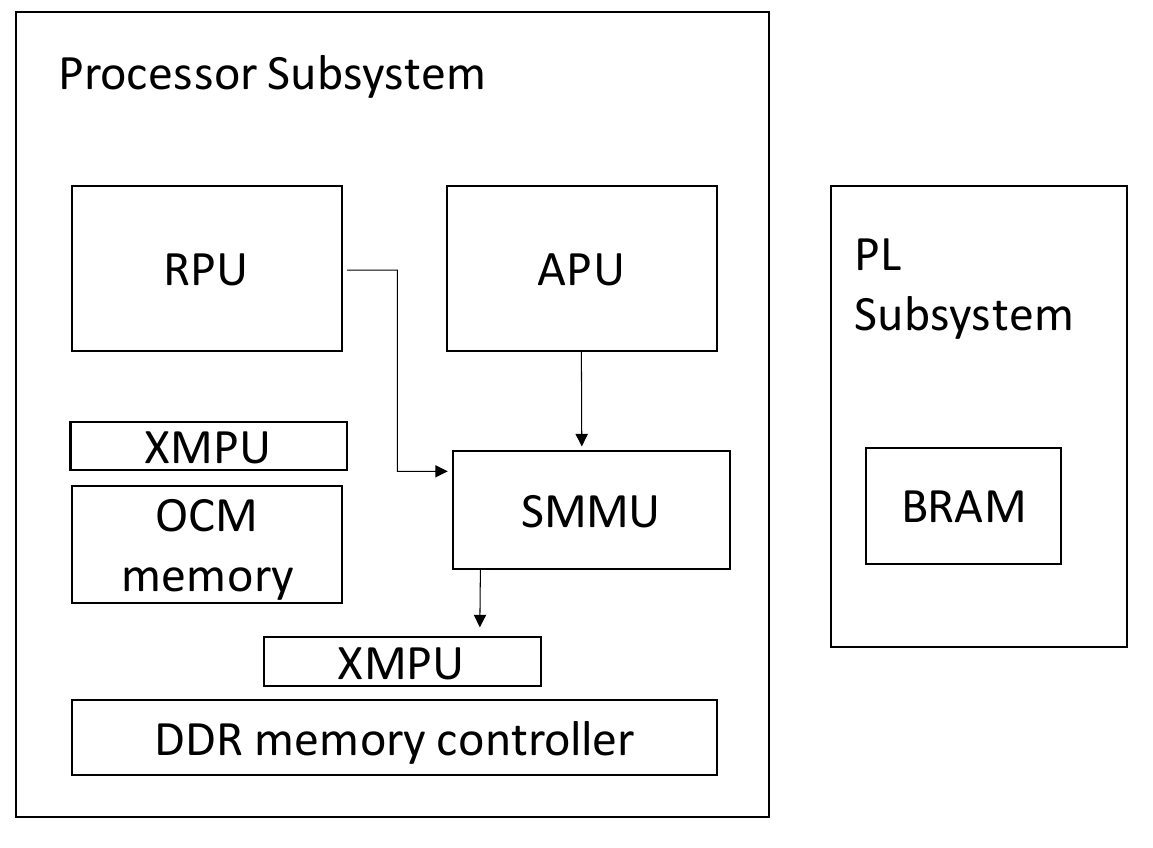}
    \caption{Zynq UltraScale+ Architecture}
    \label{fig:zynqArch}
\end{figure}

\subsection{Memory protection in Xilinx MPSoC}
In this section, we describe Xilinx Zynq UltraScale+ MPSoC hardware components that are used to isolate subsystems and protect them from each other. There are primarily two subsystems: the Processing System (PS) and the Programmable Logic (PL). And four memory regions double-data rate (DDR) memory, on-chip memory (OCM), tightly-coupled memory (TCM), and advanced eXtensible interface (AXI) block RAM in the PL system. It has eight XMPUs (Xilinx Memory Protection Units) to protect these memory regions from unauthorized access.

In this work, we focus on the  Processing Subsystem. As shown in Figure \ref{fig:zynqArch}, the PS has two main subsystems: the Application Processing Unit (APU), which is designed for general-purpose computing tasks, and the Real-time Processing Unit (RPU), which is designed for real-time applications. Xilinx protection unit is  used to create memory isolation between processes. A subsytem can configured to run in protected or unprotected modes. XMPU authorizes whether a specific process is allowed to access a specific address. Out of the eight XMPUs, six of them are used to protect the DDR memory, generating an interrupt to notify any unauthorized access.

In addition to XMPUs, the Xilinx Zynq has a  System Memory Management Unit (SMMU). SMMU extends the MMU capabilities of the processor core to the rest of the MPSoC architecture. SMMU translates virtual addresses to physical address space and can be used to restrict the reachable address space by creating multiple virtual address spaces, each with its security policy.

PetaLinux is an open-source software framework that can be used to develop and deploy embedded Linux systems on Xilinx FPGAs and SoCs.

With security blocks available Xilinx memory protection offers isolation by dividing the system memory into multiple regions, each with its own access permissions. This prevents unauthorized access to sensitive data and applications. Isolation can help to protect against unauthorized access, but it is not sufficient to secure a system. This is because memory residue from a terminated process can still be used to access sensitive data, even after the process has been isolated. Such memory may contain code, application data including sensitive information such as  passwords and encryption keys. We call this attack a \textit{Resurrection Attack}. 

\textbf{Related Attacks}: Though, not the same, a closely related attack in the programming domain is known as Use-after-Free or UAF attack \cite{yamauchi2017mitigating}. In UAF, a process, say a program written in {\sf C}, allocates memory, frees it and reallocates it using a sequence of {\sf malloc(), free()} and {\sf malloc()}. The second {\sf malloc()} inherits content from the first use without any sanitization. 
Similar attacks were mounted on GPUs by Lee \etal \cite{lee2014stealing}. They presented methods for scraping GPU memory to reconstruct information about previously running programs, revealing private information.

\textbf{Contributions}:
In this work, we show that (\textit{i}) Xilinx FPGAs do not perform automatic memory sanitization leaving memory residue, (\textit{ii}) XMPU protection mechanisms to isolate FPGA tasks or users can be altered in multiple ways to gain access to memory residue, and (\textit{iii}) present a memory scraping technique to show how sensitive information about previous programs can be reconstructed.










\section{Adversary Model}
This section provides an overview of the adversary model, encompassing both the intent and capabilities of the adversary. 

\textbf{Background}:
The Xilinx MPU is a hardware-based memory protection unit that may be used to protect memory from unauthorized access. XMPU works by assigning each process a unique memory map, which specifies which memory addresses the process can access.

\textbf{Attack question}: When an active process is terminated, does Xilinx MPU clear the memory addresses that were allocated to it? Otherwise, if this memory is allocated to a new process without clearing it first, the new process will be able to gain unauthorized access to the data that was stored in these memory addresses previously, breaking a basic goal of memory protection between processes. This question is relevant because, Xilinx allows dynamic reconfiguration and clearing the memory would add to the reconfiguration time. So how does Xilinx make tradeoff between security and performance?

\textbf{Assumptions about adversary's privileges and capabilities}: 
Our target platform is the Xilinx Zynq ZCU102 board, featuring XMPU. TThe XMPU can be configured and controlled by secure masters, such as the Platform Management Unit (PMU), Real-Time Processing Unit (RPU), or Application Processing Unit (APU) embedded in the FPGA.

The embedded operating system, PetaLinux, runs in a privileged mode that allows it to effectively program the XMPU registers through these masters. This programming capability enables PetaLinux to enforce memory isolation for specific memory addresses allocated to individual users.

In a multi-tenant scenario where multiple users share the FPGA concurrently, the need for memory protection is particularly important. Users in this environment can request memory protection from the XMPU to ensure the confidentiality and integrity of their allocated memory addresses.
    
However, once a user's session terminates, the memory addresses they were using may be available for reassignment to another user due to resource constraints. In such cases, the process of requesting memory protection and reprogramming the XMPU to provide protection for the new user's memory addresses is orchestrated through interactions with the secure masters, facilitated by PetaLinux.
    
The adversary in our study has the same privileges as any other user of the FPGA board. This means that they have the ability to request memory protection from the XMPU, use allocated memory addresses. However, the adversary's intent is different from that of regular users. The adversary is trying to exploit security vulnerabilities in the XMPU to gain unauthorized access to sensitive data of terminated process.

\section{Proposed Attack Methodology}
In this section we outline the steps need to be followed to show that memory is not initialized by a XMPU when a victim process is terminated. 

\textbf{Step 1. Polling process ID}:  
In general, any application managed by an operating system (OS) is considered a process and is allocated a unique process ID (PID). The embedded OS on the FPGA also follows this approach when dealing with applications that are offloaded onto the FPGA. The embedded OS assigns a unique ID to the application, manages resource allocation, and oversees the entire lifecycle of the process.

The adversary monitors the victim's process ID (PID) by repeatedly polling it. This monitoring continues until the victim process, which was protected by the Xilinx Memory Protection Unit (XMPU), terminates. The adversary can poll the PID using the {\sf "ps -ef" } command.

\vspace{1ex}
\textbf{Step 2. XMPU protection request to alter XMPU settings}: After the victim process terminates, the adversary requests memory allocation with protection from the embedded OS (PetaLinux). This changes the XMPU settings. However, the XMPU registers that specify the memory addresses remains unchanged. The adversary exploits this by simply requesting the activation of isolation, which gives the adversary access to the specified memory addresses.
  
\vspace{1ex}
\textbf{Step 3. Scraping virtual addresses}: At this point, the adversary does not know which physical address addresses are allocated to it by the PetaLinux system on the FPGA's onboard memory. To find out, the adversary uses the {\sf ps -ef} command to retrieve its unique process ID. With this process ID, the adversary can then access the virtual address locations where its execution is taking place. This information can be obtained by inspecting the mapping of virtual addresses allocated to the process's heap. The details of this mapping are stored in the {\sf maps} file associated with the process ID.

\vspace{1ex}
\textbf{Step 4. Mapping virtual addresses to physical addresses:} In this step, the adversary uses the page map file associated with the process ID to map the previously acquired virtual addresses to their corresponding physical addresses within the FPGA's onboard DRAM. It is important to note that this file is not accessible to users in traditional operating systems. However, the Xilinx debugger allows users to access map files. We exploit this loophole in the Xilinx debugger to obtain this information.
    
\vspace{1ex}
\textbf{Step 5. Scraping memory residue:} After obtaining the physical address, the adversary can read the content of the onboard DRAM of the FPGA. This step allows the adversary to scrape the memory residue from the terminated process. The memory residue can then be analyzed to determine what the previous process was doing and what data it was processing.

\vspace{1ex}
\textbf{Step 6. Analysis of data scraped from victim}: 
We illustrate how to analyze the data scraped from victim and gain access to information about the victim process using the following steps:
\begin{enumerate}
        \item \textit{Profiling}: The adversary first profiles the target victim processes to understand their memory layout. This involves identifying the relative memory addresses where critical data, such as signatures, vectors, or scores, is located. The adversary can use existing models or libraries in the relevant domain to perform this profiling, and the results can be saved for reference during live attacks. 
        
        \vspace{1ex}
        \item \textit{Reconstruction}: The adversary identifies the victim process by matching it to the reference file created during profiling. Once the match is confirmed, the adversary can access contents of specific memory locations where critical data is expected to reconstruct information about the previous process. The reconstruction step can be automated using software tools, which would make it even easier for the adversary to carry out the attack. However, automated reconstruction is beyond the scope of this paper.
\end{enumerate}
These steps can be used to breach the confidentiality of data that has been left unsanitized in the memory. 
 
\section{Experimental Setup}
In this section, the configuration process for the target board is outlined which is specifically tailored to facilitate the execution of the chosen attack scenario. We delve into the details of configuring the XMPU to enable and disable isolation for memory addresses relevant to the attack. This involves replicating and adapting the steps provided in the reference document offered by Xilinx \cite{XilinxVitisAI}. Furthermore, we present the specific registers that need to be programmed for the attack, ensuring comprehensive coverage of the setup process.

\begin{figure}[t!]
  \centering 
    \includegraphics[width=0.37\textwidth]{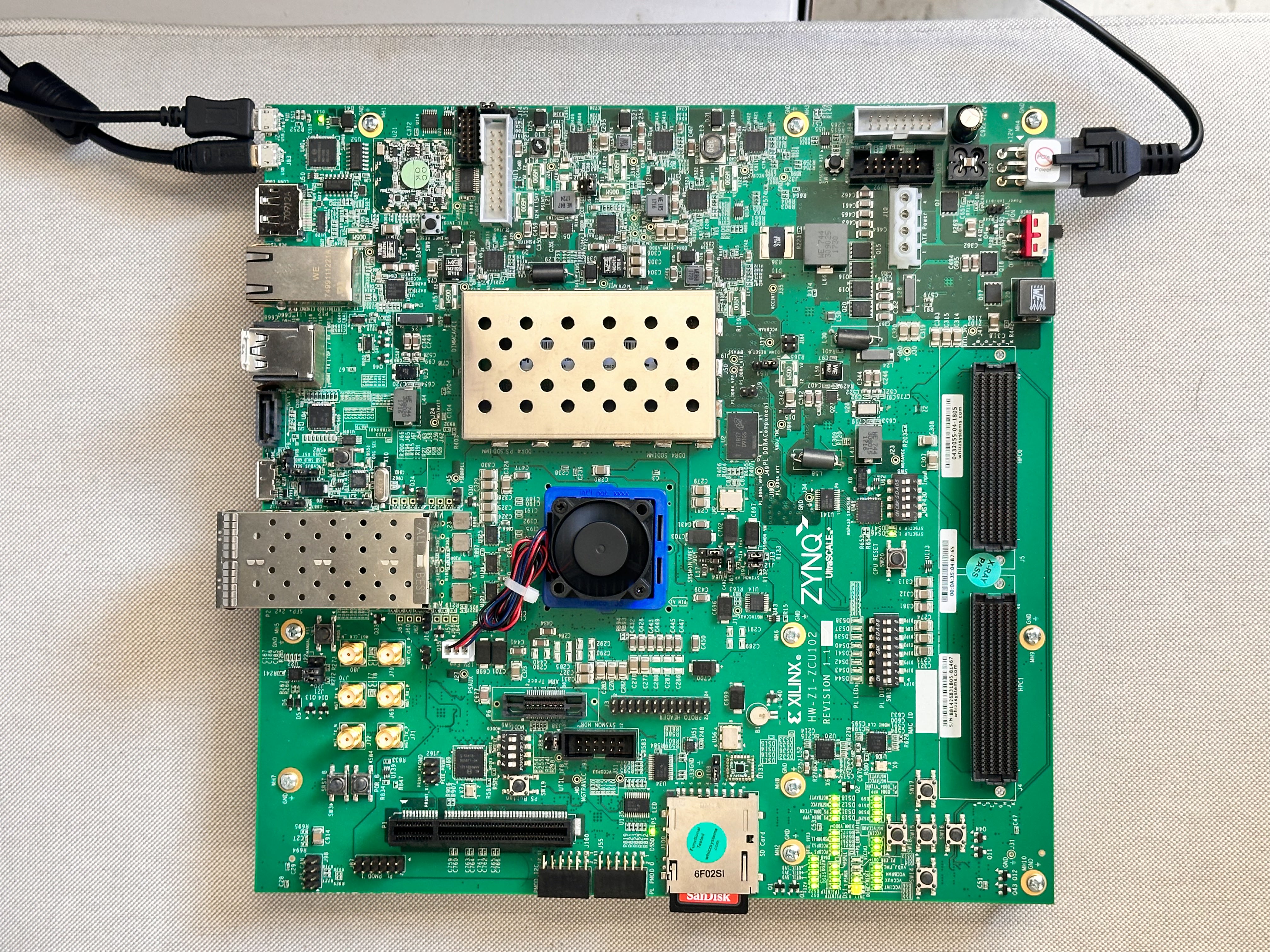} 
    \caption{Target Board (Xilinx's Zynq ZCU102)}
    \label{fig:102}
\end{figure}

\vspace{1ex}
\textbf{Setting up target board:} We conducted our experiments on the Xilinx Zynq UltraScale+ MPSOC ZCU102 Z1 Rev1.1 FPGA board as shown in Figure \ref{fig:102} . This board is based on the Zynq UltraScale+ MPSoC, which is a multiprocessing system-on-chip that includes a dual-core ARM Cortex-A72 processor, a quad-core ARM Cortex-R5F processor, and a Mali-400 MP2 graphics processing unit. It also has 1GB of DDR4 SDRAM, 128MB of QSPI flash memory, and a variety of I/O ports. We followed the step-by-step instructions from Xilinx \cite{XilinxVitisAI} to launch Peta-Linux.  PetaLinux is responsible for tasks such as assigning process PIDs, allocating resources, scheduling, and terminating processes.

\begin{enumerate}
    \item The process begins by flashing the OS image provided by Xilinx for the ZCU102 to an SD card. This image contains the operating system (PetaLinux) and software components necessary for the FPGA board. Once the OS image was successfully flashed to the SD card, we booted the board by inserting the SD card and turning it on.
    
    \item Once the board is booted, we establish a remote connection to it using the Ethernet interface. This allows us to communicate and interact with the board from a remote location.
    
    \item Finally, we install the Vitis AI runtime on the target board. This runtime environment includes a collection of example machine learning models from various vendors, providing a comprehensive set of pre-built models for testing and experimentation.
\end{enumerate}

The above steps create a secure environment that allows us to configure XMPU security and explore its security vulnerabilities.

\vspace{1ex}
\subsection{Attack sequence involving alteration of XMPU settings} 
The attack sequence for demonstrating lack of XMPU protection for a terminated process are derived from modifications to the steps outlined in the "APU Fault Injection Software Application Project" \cite{XilinxXapp1320}. The original intent of this project is to showcase XMPU's protection capabilities for an active process. We will now provide an overview of attack sequence with the modified procedure below.

\begin{itemize}
    \item\textbf{Victim Process}: The RPU runs an application on behalf of a process (victim), using specific memory addresses in the DRAM that are protected by the XMPU to prevent unauthorized access. The threat of unauthorized access comes from an adversary running on the APU. To demonstrate that the XMPU does not properly clear memory addresses designated for protection after a process finishes, we do the following:

\vspace{1ex}
\begin{enumerate}
    \item We create a APU application (proxy for victim) to write keywords into memory addresses designated for use by the RPU. These addresses  were intended to be secure when the RPU accessed them.
        
    \item To run the application on RPU, we then asked PetaLinux to activate the XMPU's isolation mechanism. This entailed programming the XMPU with the relevant memory addresses  and enabling isolation.
        
    \item Once activated, this application writes specific keywords to these memory locations. At this point, the adversary on the APU attempted to access the aforementioned memory addresses . The XMPU immediately prevented this access by generating an interrupt, thus upholding the security measures set by the XMPU.
\end{enumerate}

\vspace{1ex}
\item \textbf{Adversary Process}: We describe two scenarios for the attack process with and without XMPU protection.

\vspace{1ex}
\begin{enumerate} 
    \item \textit{XMPU disabled}: After termination of the victim process, we disable the XMPU protection for this memory addresses as the process is no longer alive.
    
    We then ran an adversary process to read the memory residue. We found that the adversary was able to read the keywords written by the victim process and the keywords written by the adversary, which were written before the RPU initiated a transaction.  Even though the victim process had terminated and the XMPU was no longer active, the victim's data persisted within these memory locations allowing a different process to read it.\vspace{1ex}

    \item \textit{XMPU enabled}: In this scenario, when the victim's process is terminated, we request memory protection for executing the adversary process. The embedded OS allocated the same memory locations (as they remained unchanged in the XMPU registers) to the adversary process. Once again, our adversary process gained access to the victim's data and its own data (written prior to RPU's initiation), revealing a lack of proper memory initialization by the XMPU after a process has terminated.
\end{enumerate}
\end{itemize}



 \begin{figure}[t!]
   \centering 
     \includegraphics[width=0.37\textwidth]{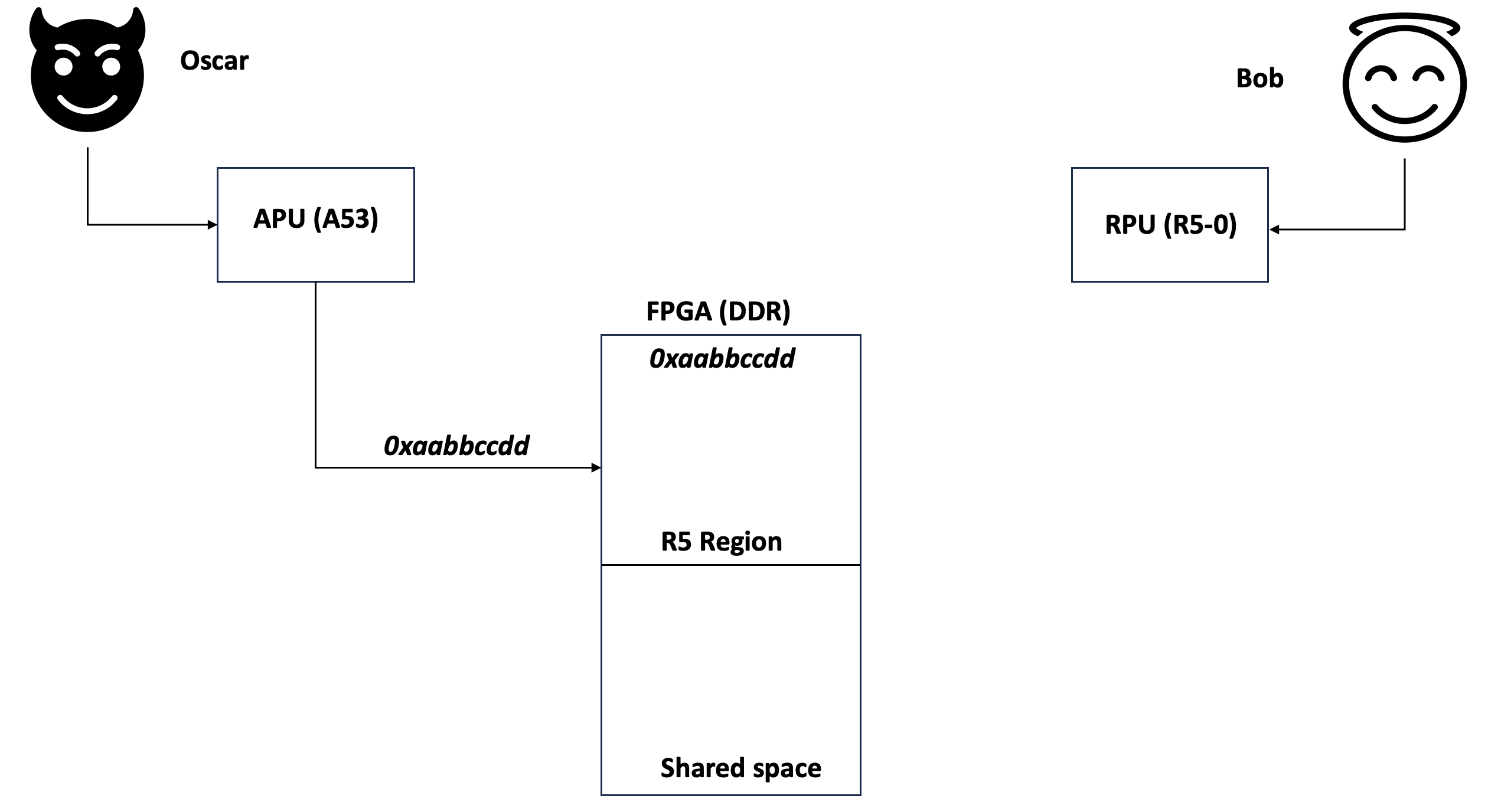} 
     \caption{Oscar writes into Bob's memory addresses when XMPU is disabled.}
     \label{fig:1}
 \end{figure}

 \begin{figure}[t!]
   \centering 
     \includegraphics[width=0.37\textwidth]{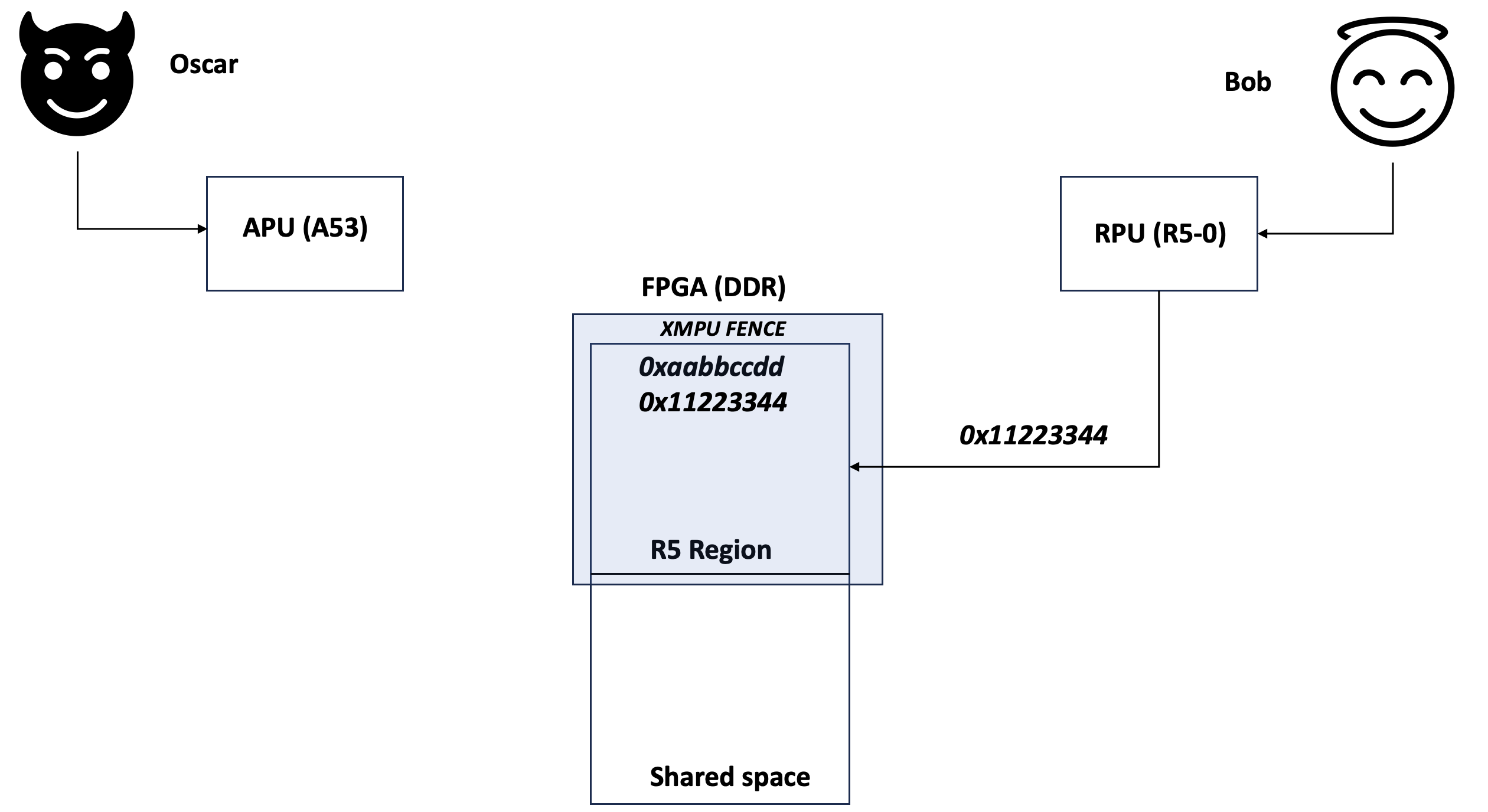} 
     \caption{Bob enables XMPU fence and writes his data.}
     \label{fig:2}
 \end{figure}

 \begin{figure}[t!]
   \centering 
     \includegraphics[width=0.37\textwidth]{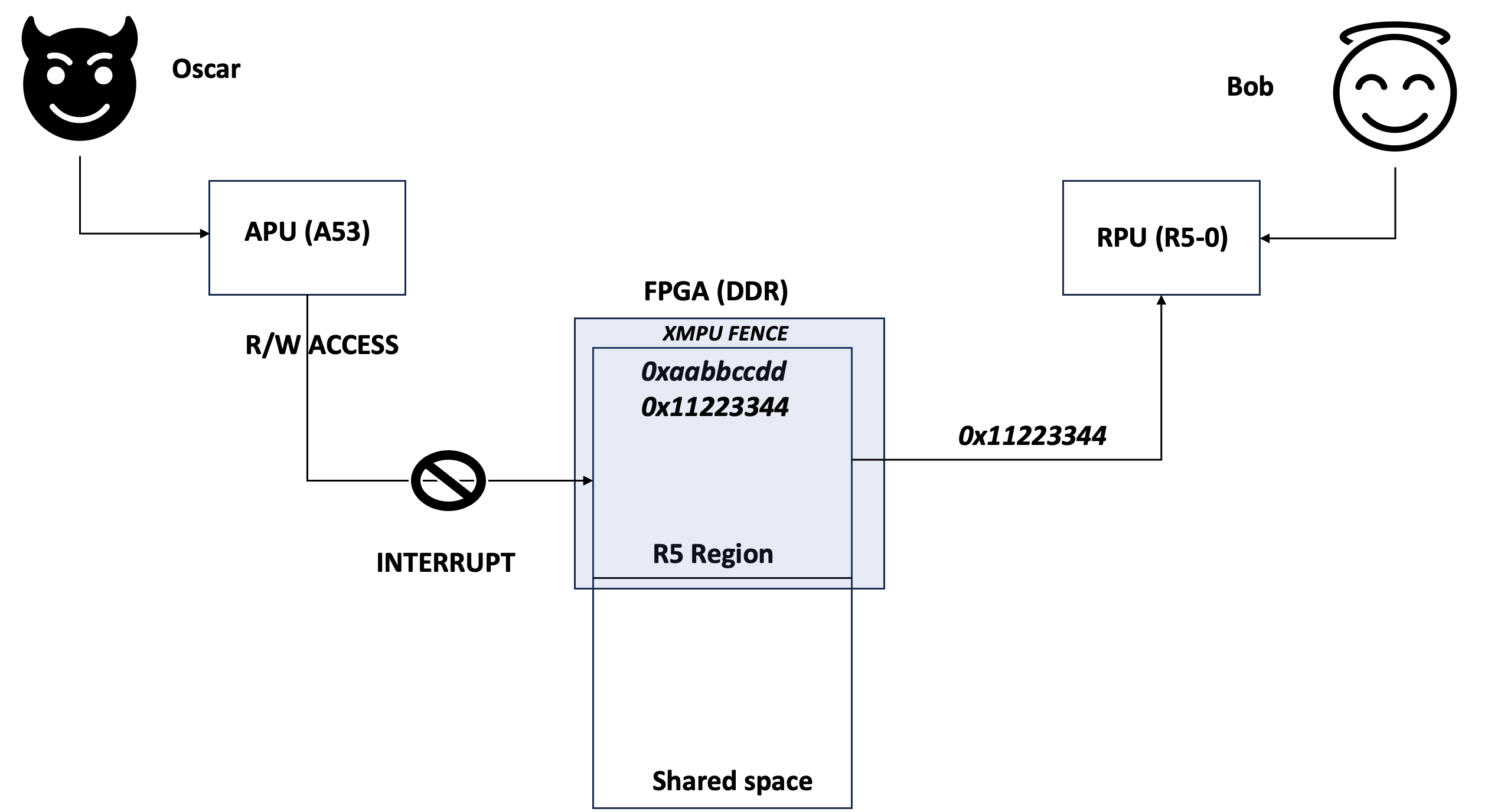} 
     \caption{Oscar's access attempt to Bob's addresses  is denied, triggering an interrupt.}
     \label{fig:3}
 \end{figure}

 \begin{figure}[t!]
   \centering 
     \includegraphics[width=0.37\textwidth]{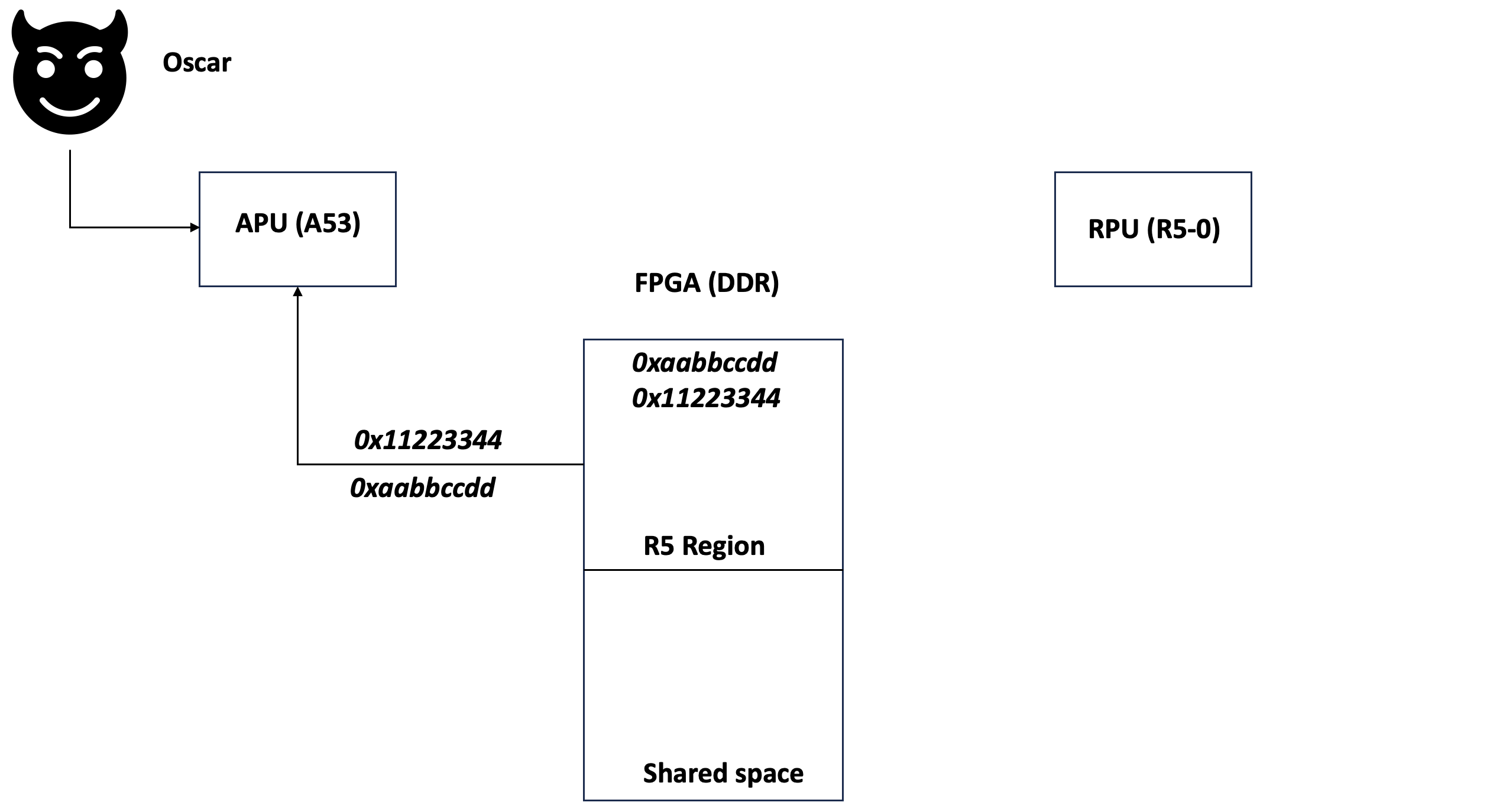} 
     \caption{After Bob's termination, the XMPU fence is lifted, granting Oscar access to Bob's memory contents.}
     \label{fig:4}
 \end{figure}

\begin{figure}[t!]
  \centering 
    \includegraphics[width=0.35\textwidth]{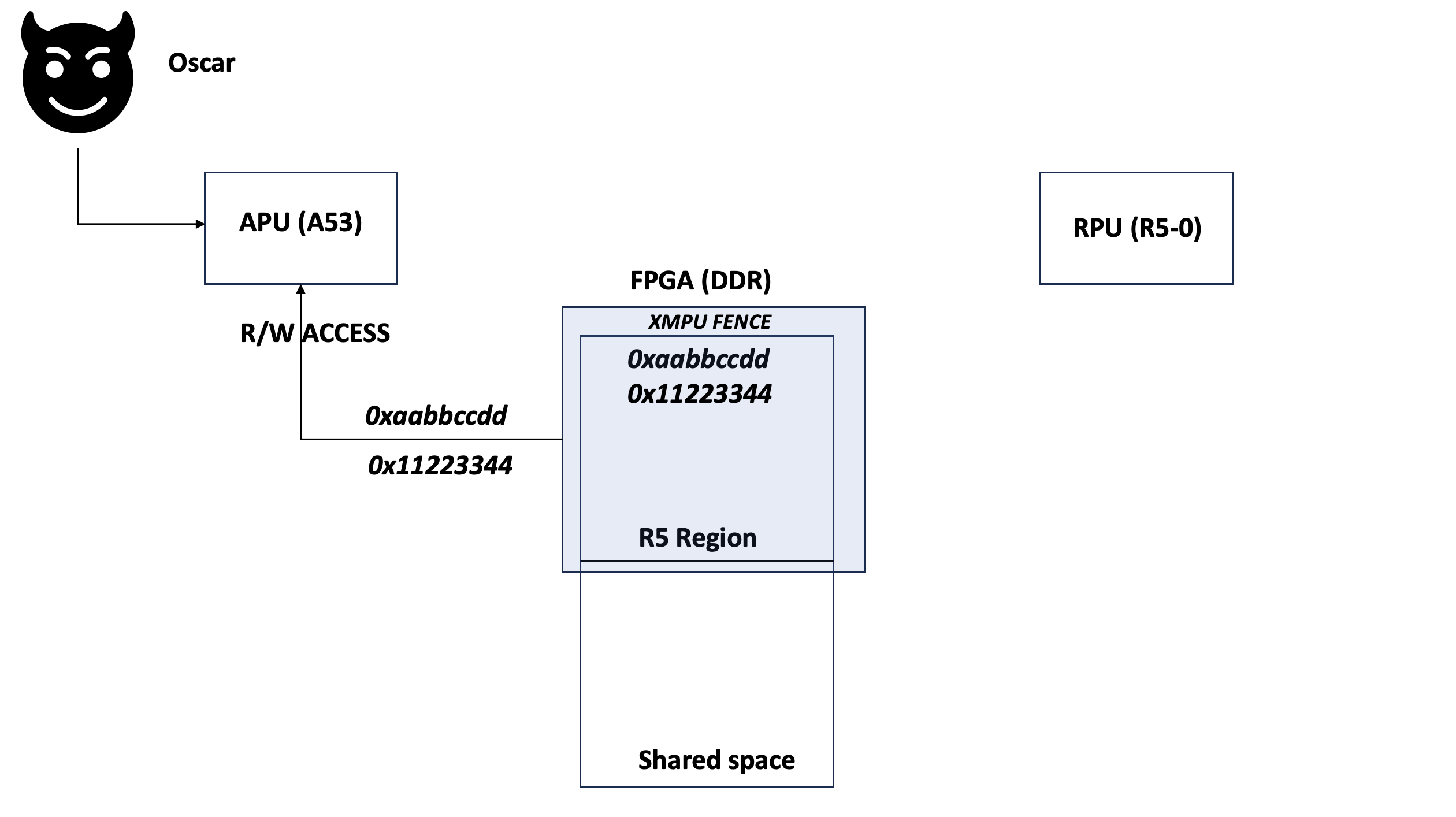} 
    \caption{Following Bob's termination, Oscar initiates a request for isolation from XMPU, thereby gaining access to the contents of Bob's memory.}
    \label{fig:5}
\end{figure}

\vspace{1ex}
Figures \ref{fig:1} to \ref{fig:5} illustrate the overview of attack procedure described above. Oscar represents the adversary using the APU, while Bob represents the victim using the RPU. 
In Figure \ref{fig:1}, Oscar exploits the disabled XMPU to write data into Bob's memory addresses. 
In Figure \ref{fig:2}, Bob activates the XMPU fence and adds his data. 
In Figure \ref{fig:3}, Oscar's attempt to access Bob's addresses  is thwarted, resulting in an interrupt. 
In Figure \ref{fig:4}, After Bob's process ends, the XMPU fence is lifted, allowing Oscar to access the contents within Bob's memory. 
In Figure \ref{fig:5}, In an alternate scenario, the XMPU isolation remains active after Bob's termination. Oscar then initiates a request for isolation from the XMPU, which leads to the same memory addresses  being accessed but with the ID in the XMPU configuration changed from Bob's to Oscar's.

The above sequence highlights the role of the XMPU in governing memory access when Bob's process is active, while also highlighting the potential exposure of Bob's data following his process termination.

\vspace{1ex}
\subsection{Programming steps for executing the attack sequence}
We illustrate the programming steps for demonstrating the lack of XMPU protection for a terminated process/application ran on the RPU. The RPU can access various types of memory, namely OCM, DDR, and ATCM. For the purpose of protection,  memory regions are designated into three classes: secure, non-secure, or not-defined. The designations are explained below.

\begin{itemize}
    \item \textit{Secure (S):} This class of memory is accessible only to application running on RPU. The data stored in this memory is considered to be confidential and should not be accessible to any other process. These specific addresses are configured in the XMPU register with an ID that permits access only to the RPU application. Any attempt of unauthorized access triggers an error interrupt.
    \vspace{1ex}
    \item \textit{Non-Secure (NS):} This class of memory is accessible by both the task running on the RPU and tasks running on other IPs, such as the APU. The data stored in this memory is less critical and is intended to be shared among multiple IPs. In this scenario, the XMPU is configured with these memory addresses, but isolation is not enabled for them.
    \vspace{1ex}
    \item \textit{Not-Defined (ND):} The Not-defined class comprises memory addresses located within various memory regions that are currently unallocated and available for use by applications or IPs as needed. When a request is made for access to these addresses, their configuration in the XMPU can be determined based on the criticality of the user's data. They can be set as either Secure or Non-Secure addresses accordingly.
\end{itemize}

\vspace{1ex}
\textbf{XMPU Register Configuration}:
The following XMPU registers are used to configure various memory regions as secure or non-secure:
\begin{itemize}
    \item \textit{Control registers}: These registers serve the purpose of enabling or disabling read and write access to the designated memory addresses as configured in the associated configuration registers. They are designed with a width of 32 bits, and their initial default value is set to 0x00000013. This default value implies that any user can access the memory regions without any isolation. To implement isolation and restrict access, these registers need to be programmed with the value 0x00000010.
    \vspace{1ex}
    \item \textit{SMID register}: The System Management ID (SMID) register is a 32-bit wide register, with the least significant 10 bits designated for configuration with the intended SMIDs required for accessing the respective addresses. If any other ID attempts to access these addresses, it triggers an interrupt, initiating an error handling process within the system. Additionally, the ID values can be masked with a 10-bit mask value to make them invisible, thereby preventing potential side-channel attacks on IDs. 
    \vspace{1ex}
    \item \textit{END\_HI/LOW and START\_HI/LOW registers}: The combination of the {\sf START\_HI} and {\sf END\_HI} registers, when used together as a pair, constitute a 44-bit start address within a memory region. In this pair, {\sf START\_HI} uses its least significant 12 bits, while {\sf END\_HI} employs its entire 32 bits to create a 44-bit start address. Similarly, for the end address, it is formed by combining {\sf END\_LOW[11:0]} with {\sf END\_HI[31:0]}. The classification of these start and end addresses as either "secure" or "non-secure" is determined by the programming performed in the control registers, followed by the configuration of the corresponding ID that is permitted in the System Management ID (SMID) register.
    \vspace{1ex}
    \item \textit{LOCK registers}: When enabled (zero\textsuperscript{th} bit of the register), all XMPU registers can only be programmed once (during boot), and their settings can only be reset through an internal or external Power On Reset (POR). In scenarios involving multiple tenants, the lock register should be set to a value of 0x0. This allows for the flexible configuration of memory addresses  within the XMPU registers in response to specific demands. It also facilitates the activation or deactivation of isolation as required.
\end{itemize}

\vspace{1ex}
XMPU registers are configured with memory addresses based on the criticality and necessity of data handled by a victim process, as well as the availability of addresses in the memory regions. These configurations classify memory regions as either "secure" or "non-secure" depending on their specific usage requirements. This flexible approach ensures that data security is tailored to the unique demands of each process while optimizing resource allocation.

Once the ZCU 102 board is successfully configured, we execute the attack sequence which allows us to investigate XMPU memory residue protection. We find that, Xilinx currently does not sanitize the memory after a process is terminated, which can leave memory residue with sensitive data.
\section{Results}
In this section, we provide  results with illustrations from each step described in Section 3. These results also highlight the implementation aspects of the attack.

\vspace{1ex}
\textbf{Watermarking of RPU memory region by APU: } As described in the attack scenario, the APU, acting as a proxy for the victim, writes specified keyword {\sf 0x11223344} into specific memory locations of the RPU. These memory locations are then programmed into the specialized XMPU registers to create isolation during RPU usage. These APU write operations demonstrate that the XMPU does not initialize memory addresses for isolation before or after the start and end of a process, as described next.

\begin{figure}[t!]
\begin{verbbox}[\scriptsize\slshape]
 Disabling XMPU

         Disabling ...
                 Writing DDR_XMPU0_CTRL                 ...  PASS!
                 Writing DDR_XMPU1_CTRL                 ...  PASS!
                 Writing DDR_XMPU2_CTRL                 ...  PASS!
                 Writing DDR_XMPU3_CTRL                 ...  PASS!

   Read/Write Of Various Memories

         Read/Write Of RPU(Secure) Memory
                 Reading RPU_OCM_S_BASE                 ...  PASS!
                 Writing RPU_OCM_S_BASE                 ...  PASS!
                 Reading RPU_DDR_LOW_S_BASE             ...  PASS!
                 Writing RPU_DDR_LOW_S_BASE             ...  PASS!
                 Reading RPU_ATCM_S_BASE                ...  PASS!
                 Writing RPU_ATCM_S_BASE                ...  PASS!

    APU has written 0x11223344 into RPU(Secure) Memory

\end{verbbox}
\resizebox{8.6cm}{4.5cm}{\fbox{\theverbbox}}
\caption{APU (adversary) writes data into secure memory addresses of RPU.}
\label{fig:a1}
\end{figure}

\begin{figure}[t!]
\begin{verbbox}[\scriptsize\slshape]
 
    Read/Write Of Various Memories

         Read/Write Of RPU(S) Memory
                 Reading RPU_OCM_S_BASE                 ...  FAILED!
                 Writing RPU_OCM_S_BASE                 ...  FAILED!
                 Reading RPU_DDR_LOW_S_BASE             ...  FAILED!
                 Writing RPU_DDR_LOW_S_BASE             ...  FAILED!
                 Reading RPU_ATCM_S_BASE                ...  FAILED!
                 Writing RPU_ATCM_S_BASE                ...  FAILED!
          
    APU is denied access into RPU(Secure) Memory

\end{verbbox}
\resizebox{8.6cm}{2.5cm}{\fbox{\theverbbox}}
\caption{APU (adversary) is denied access to the secure memory addresses of RPU.}
\label{fig:a2}
\end{figure}

Figure \ref{fig:a1} shows how to disable isolation protections. First, the APU writes values to the XMPU's CTRL registers to disable isolation. These writes are valid transactions (denoted by "PASS!"), which prevents interrupts from being triggered by the error handler. Next, the APU writes data to different memories, including the On-chip Memory (OCM) used by secure masters, DDR, which is the on-board memory used widely across FPGA IPs, and Tightly Coupled Memory (ATCM). The RPU frequently accesses these memories from FPGA reset through shutdown.

\vspace{1ex}
\textbf{XMPU protection activation and RPU watermarking:} Then, the RPU process (victim) enables isolation protections and stores the keyword {\sf 0xaabbccdd} in its memory addresses at locations that the APU had not written to previously. When the APU attempts to access these protected addresses, it is blocked and an interrupt is raised, which is handled by the error handler. This demonstrates that the XMPU provides isolation for an active process.

Although the RPU enabling the XMPU and accessing its own addresses is similar to what is shown in Figure \ref{fig:a1}, Figure \ref{fig:a2} shows that the APU is now denied access to the RPU's protected memory addresses.

\vspace{1ex}
\textbf{XMPU protection deactivation and watermark check:} After the victim process is terminated and XMPU isolation is deactivated, the APU can still access the memory locations where it originally wrote the value {\sf 0x11223344}. It can also access the memory locations where the RPU had written the value "0xaabbccdd". This is shown in Figure \ref{fig:a3}. This shows that XMPU does not clear memory after a process terminates, and it does not clear memory even when isolation is disabled.

\begin{figure}[t!]
\begin{verbbox}[\scriptsize\slshape]
 Disabling XMPU

         Enabling ...
                 Writing DDR_XMPU0_CTRL                 ...  PASS!
                 Writing DDR_XMPU1_CTRL                 ...  PASS!
                 Writing DDR_XMPU2_CTRL                 ...  PASS!
                 Writing DDR_XMPU3_CTRL                 ...  PASS!

   Read/Write Of Various Memories

         Read/Write Of RPU(Secure) Memory
                 Reading RPU_OCM_S_BASE                 ...  PASS!
                 Writing RPU_OCM_S_BASE                 ...  PASS!
                 Reading RPU_DDR_LOW_S_BASE             ...  PASS!
                 Writing RPU_DDR_LOW_S_BASE             ...  PASS!
                 Reading RPU_ATCM_S_BASE                ...  PASS!
                 Writing RPU_ATCM_S_BASE                ...  PASS!

    APU has read 0x11223344, 0xaabbccdd from RPU(Secure) Memory

\end{verbbox}
\resizebox{8.6cm}{4.5cm}{\fbox{\theverbbox}}
\caption{APU (Adversary) Accesses Both Its Data and the Victim's Data from RPU's Secure Memory addresses.}
\label{fig:a3}
\vspace{-2ex}
\end{figure}

\vspace{1ex}
These experiments were conducted in a controlled environment, where we have full knowledge of the specific memory addresses that we wanted to manipulate. These addresses were all associated with the RPU subsystem. We also had the flexibility to program the XMPU, enabling or disabling it as needed for our attack scenario.

This simulated setup emulated a multi-tenant environment, which is a real-world scenario where adversarial applications may try to infiltrate the memory addresses of a victim application running on the RPU. We did not just focus on RPU memory, but also extended our analysis to include shared memory space between the APU and RPU, and various peripheral components. 

Figure \ref{fig:a4} illustrates our findings based on the ZCU102 platform. This figure shows that the XMPU can effectively isolate active processes. However, it also reveals a critical vulnerability: the inability to sanitize memory after a process terminates or even when isolation is reset (\textit{i.e.}, transitioning from an enabled to a disabled state).

\begin{figure}[t!]
\begin{verbbox}[\scriptsize\slshape]
 ---Starting Fault Injection Test (Running on the APU)---

   (S)=Secure, (NS)=Non-Secure, (ND)=Not-Defined
   Memories
        RPU_OCM_S_BASE                : OCM Secure Memory Base Address in RPU Sub-System
        RPU_OCM_NS_SHARED_BASE        : OCM Non-Secure Memory Base Address in both RPU and APU Sub-Systems
        RPU_ATCM_S_BASE               : R5 TCM Bank A Secure Memory Base Address in RPU Sub-System
        RPU_DDR_LOW_S_BASE            : DDR Secure Memory Base Address in RPU Sub-System
        RPU_DDR_LOW_NS_SHARED_BASE    : DDR Non-Secure Memory Base Address in both RPU and APU Sub-Systems
        APU_OCM_NS_SHARED_BASE        : OCM Non-Secure Memory Base Address in both APU and RPU Sub-Systems
        APU_DDR_LOW_NS_BASE           : DDR Non-Secure Memory Base Address in APU Sub-System
        APU_DDR_LOW_NS_SHARED_BASE    : DDR Non-Secure Memory Base Address in Both APU and RPU Sub-Systems
        UNDEFINED_DDR_MEMORY_BASE     : Memory Base Address Not Defined in Any Sub-System Peripherals
        APU_UART0_NS_START            : Non-Secure UART0 in APU Sub-System
        APU_SWDT0_NS_START            : Non-Secure UART0 in APU Sub-System
        APU_TTC0_NS_START             : Non-Secure UART0 in APU Sub-System
        APU_UART0_NS_START            : Non-Secure UART0 in APU Sub-System
        SHARED_GPIO_NS_START          : Shared Non-Secure GPIO
        RPU_UART1_S_START             : Secure UART1 in RPU Sub-System
        RPU_SWDT1_S_START             : Secure SWDT1 in RPU Sub-System
        RPU_TTC1_S_START              : Secure TTC1 in RPU Sub-System
        RPU_I2C1_S_START              : Secure I2C1 in RPU Sub-System
        DDR_XMPU0/1/2/3_CTRL          : Enables and Disables Security isolation
*******************************************
 Disabling XMPU
         Disabling ...
                 Writing DDR_XMPU0_CTRL                 ...  PASS!
                 Writing DDR_XMPU1_CTRL                 ...  PASS!
                 Writing DDR_XMPU2_CTRL                 ...  PASS!
                 Writing DDR_XMPU3_CTRL                 ...  PASS!
 Read/Write Of Various Memories
         Read/Write Of RPU(S) Memory
                 Reading RPU_OCM_S_BASE                 ...  PASS!
                 Writing RPU_OCM_S_BASE                 ...  PASS!
                 Reading RPU_DDR_LOW_S_BASE             ...  PASS!
                 Writing RPU_DDR_LOW_S_BASE             ...  PASS!
                 Reading RPU_ATCM_S_BASE                ...  PASS!
                 Writing RPU_ATCM_S_BASE                ...  PASS!
         Read/Write Of RPU(NS) Memory
                 Reading RPU_OCM_NS_SHARED_BASE         ...  PASS!
                 Writing RPU_OCM_NS_SHARED_BASE         ...  PASS!
                 Reading RPU_DDR_LOW_NS_SHARED_BASE     ...  PASS!
                 Writing RPU_DDR_LOW_NS_SHARED_BASE     ...  PASS!
         Read/Write Of APU(S) Memory
                ---This combination does not exist
                ---APU has no (S)ecure memory
         Read/Write Of APU(NS)
                 Reading APU_OCM_NS_SHARED_BASE         ...  PASS!
                 Writing APU_OCM_NS_SHARED_BASE         ...  PASS!
                 Reading APU_DDR_LOW_NS_BASE            ...  PASS!
                 Writing APU_DDR_LOW_NS_BASE            ...  Skipped to avoid memory collision!
                 Reading APU_DDR_LOW_NS_SHARED_BASE     ...  PASS!
                 Writing APU_DDR_LOW_NS_SHARED_BASE     ...  PASS!
         Read/Write Of Undefined (ND) Memory
                 Reading UNDEFINED_DDR_MEMORY_BASE      ...  PASS!
                 Writing UNDEFINED_DDR_MEMORY_BASE      ...  PASS!
   Reading Sub-System Peripherals
         APU Peripherals
                 Reading APU_UART0_NS_START             ...  PASS!
                 Reading APU_SWDT0_NS_START             ...  PASS!
                 Reading APU_TTC0_NS_START              ...  PASS!
         RPU Peripherals
                 Reading RPU_UART1_S_START              ...  PASS!
                 Reading RPU_SWDT1_S_START              ...  PASS!
                 Reading RPU_TTC1_S_START               ...  PASS!
                 Reading RPU_I2C1_S_START               ...  PASS!
         Shared Peripherals
                 Reading SHARED_GPIO_NS_START           ...  PASS!

APU has written 0x11223344 into DRAM
*******************************************
 APU tries to disable XMPU which is now programmed by RPU
         Disabling ...
                 Writing DDR_XMPU0_CTRL                 ...  FAILED!
                 Writing DDR_XMPU1_CTRL                 ...  FAILED!
                 Writing DDR_XMPU2_CTRL                 ...  FAILED!
                 Writing DDR_XMPU3_CTRL                 ...  FAILED!
   Read/Write Of Various Memories
         Read/Write Of RPU(S) Memory
                 Reading RPU_OCM_S_BASE                 ...  FAILED!
                 Writing RPU_OCM_S_BASE                 ...  FAILED!
                 Reading RPU_DDR_LOW_S_BASE             ...  FAILED!
                 Writing RPU_DDR_LOW_S_BASE             ...  FAILED!
                 Reading RPU_ATCM_S_BASE                ...  FAILED!
                 Writing RPU_ATCM_S_BASE                ...  FAILED!
         Read/Write Of RPU(NS) Memory
                 Reading RPU_OCM_NS_SHARED_BASE         ...  PASS!
                 Writing RPU_OCM_NS_SHARED_BASE         ...  PASS!
                 Reading RPU_DDR_LOW_NS_SHARED_BASE     ...  PASS!
                 Writing RPU_DDR_LOW_NS_SHARED_BASE     ...  PASS!
         Read/Write Of APU(S) Memory
                ---This combination does not exist
                ---APU has no (S)ecure memory
         Read/Write Of APU(NS)
                 Reading APU_OCM_NS_SHARED_BASE         ...  PASS!
                 Writing APU_OCM_NS_SHARED_BASE         ...  PASS!
                 Reading APU_DDR_LOW_NS_BASE            ...  PASS!
                 Writing APU_DDR_LOW_NS_BASE            ...  Skipped to avoid memory collision!
                 Reading APU_DDR_LOW_NS_SHARED_BASE     ...  PASS!
                 Writing APU_DDR_LOW_NS_SHARED_BASE     ...  PASS!
         Read/Write Of Undefined (ND) Memory
                 Reading UNDEFINED_DDR_MEMORY_BASE      ...  PASS!
                 Writing UNDEFINED_DDR_MEMORY_BASE      ...  PASS!
   Reading Sub-System Peripherals
         APU Peripherals
                 Reading APU_UART0_NS_START             ...  PASS!
                 Reading APU_SWDT0_NS_START             ...  PASS!
                 Reading APU_TTC0_NS_START              ...  PASS!
         RPU Peripherals
                 Reading RPU_UART1_S_START              ...  FAILED!
                 Reading RPU_SWDT1_S_START              ...  FAILED!
                 Reading RPU_TTC1_S_START               ...  FAILED!
                 Reading RPU_I2C1_S_START               ...  FAILED!
         Shared Peripherals
                 Reading SHARED_GPIO_NS_START           ...  PASS!
 APU cannot read  DRAM
*******************************************
 APU disabling XMPU (Victim on RPU is terminated)
         Disabling ...
                 Writing DDR_XMPU0_CTRL                 ...  PASS!
                 Writing DDR_XMPU1_CTRL                 ...  PASS!
                 Writing DDR_XMPU2_CTRL                 ...  PASS!
                 Writing DDR_XMPU3_CTRL                 ...  PASS!
  Read/Write Of Various Memories
         Read/Write Of RPU(S) Memory
                 Reading RPU_OCM_S_BASE                 ...  PASS!
                 Writing RPU_OCM_S_BASE                 ...  PASS!
                 Reading RPU_DDR_LOW_S_BASE             ...  PASS!
                 Writing RPU_DDR_LOW_S_BASE             ...  PASS!
                 Reading RPU_ATCM_S_BASE                ...  PASS!
                 Writing RPU_ATCM_S_BASE                ...  PASS!
         Read/Write Of RPU(NS) Memory
                 Reading RPU_OCM_NS_SHARED_BASE         ...  PASS!
                 Writing RPU_OCM_NS_SHARED_BASE         ...  PASS!
                 Reading RPU_DDR_LOW_NS_SHARED_BASE     ...  PASS!
                 Writing RPU_DDR_LOW_NS_SHARED_BASE     ...  PASS!
         Read/Write Of APU(S) Memory
                ---This combination does not exist
                ---APU has no (S)ecure memory
         Read/Write Of APU(NS)
                 Reading APU_OCM_NS_SHARED_BASE         ...  PASS!
                 Writing APU_OCM_NS_SHARED_BASE         ...  PASS!
                 Reading APU_DDR_LOW_NS_BASE            ...  PASS!
                 Writing APU_DDR_LOW_NS_BASE            ...  Skipped to avoid memory collision!
                 Reading APU_DDR_LOW_NS_SHARED_BASE     ...  PASS!
                 Writing APU_DDR_LOW_NS_SHARED_BASE     ...  PASS!
 Read/Write Of Undefined (ND) Memory
                 Reading UNDEFINED_DDR_MEMORY_BASE      ...  PASS!
                 Writing UNDEFINED_DDR_MEMORY_BASE      ...  PASS!
   Reading Sub-System Peripherals
         APU Peripherals
                 Reading APU_UART0_NS_START             ...  PASS!
                 Reading APU_SWDT0_NS_START             ...  PASS!
                 Reading APU_TTC0_NS_START              ...  PASS!
         RPU Peripherals
                 Reading RPU_UART1_S_START              ...  PASS!
                 Reading RPU_SWDT1_S_START              ...  PASS!
                 Reading RPU_TTC1_S_START               ...  PASS!
                 Reading RPU_I2C1_S_START               ...  PASS!
         Shared Peripherals
                 Reading SHARED_GPIO_NS_START           ...  PASS!
 APU has read 0x11223344 and 0xaabbccdd from DRAM
---Fault Injection Test Complete---

\end{verbbox}
\resizebox{8.6cm}{21.5cm}{\fbox{\theverbbox}}
\caption{Extensive Attack Scenario by APU on RPU.}
\label{fig:a4}
\end{figure}

The results shown above indicate that an adversary can enter as a new process after XMPU's isolation is disabled, and then read the terminated victim's data. This suggests that XMPU does not sanitize memory after a process has been terminated, leaving it vulnerable to readout. The same results were observed when XMPU remained enabled as the adversary entered as a new process. In this case, instead of disabling and then re-enabling isolation for a new process, the XMPU configuration register was simply updated with a new ID to allow access. However, the memory addresses and isolation settings remained unchanged in the XMPU registers. As a result, the adversary process was still able to read these previously accessed memory locations from the terminated process, highlighting that XMPU did not initialize memory, consistent with the results shown above.

Next, we will illustrate how the adversary can access the locations of the previous process and attempt to determine its nature. This will be demonstrated through a scenario in which the victim executes a critical application named {\sf critical\_application.py} which only writes 0xffffffff into memory, and we will show how the adversary can access this data and leverage this information. The adversary runs an application named {\sf adversary.py} which only scrapes data from the physical addresses obtained in the steps below.

\vspace{1ex}
\noindent\textbf{Step 1. Polling for process ID}: Figures \ref{fig:a5} and \ref{fig:a6} show the active processes (pids) obtained from the attacker's terminal by executing the {\sf ps -ef} command. Figure \ref{fig:a5} shows the processes running {\sf critical\_application.py}, while Figure \ref{fig:a6} shows the processes after it is terminated. 

After the victim is terminated, the adversary requests for its memory protection by isolation (Step2) and start execution. Figure \ref{fig:a7} shows that the adversary starts executing its own code with pid {\sf 2840}.

\begin{figure}[t!]
\begin{verbbox}[\scriptsize\slshape]
1874       1  0 Jul12 ?        00:00:00 /usr/sbin/dropbear -i -r /etc/dropbear/dropbear_rsa_host_key -B
1875    1874  0 Jul12 pts/0    00:00:00 -sh
2056       2  0 Jul12 ?        00:00:00 [kworker/3:2-cgroup_destroy]
2429       1  0 05:11 ?        00:00:00 /usr/sbin/dropbear -i -r /etc/dropbear/dropbear_rsa_host_key -B
2430    2429  0 05:11 pts/1    00:00:00 -sh
2456       2  0 05:14 ?        00:00:00 [kworker/2:0H]
2457       2  0 05:14 ?        00:00:00 [DPUCZDX8G_2]
2458       2  0 05:14 ?        00:00:00 [DPUCZDX8G_1]
2776       2  0 11:39 ?        00:00:00 [kworker/u8:1-events_unbound]
2802       2  0 11:55 ?        00:00:00 [kworker/3:1H]
2803       2  0 11:55 ?        00:00:00 [kworker/1:2H]
2811       2  0 12:04 ?        00:00:00 [kworker/0:2-events_power_efficient]
2819       2  0 12:16 ?        00:00:00 [kworker/u8:0-events_unbound]
2828       2  0 12:27 ?        00:00:00 [kworker/0:1-mm_percpu_wq]
2829       2  0 12:29 ?        00:00:00 [kworker/u8:2-events_unbound]
2831     849  0 12:30 ?        00:00:00 systemd-userwork
2832     849  0 12:30 ?        00:00:00 systemd-userwork
2833     849  0 12:30 ?        00:00:00 systemd-userwork
2834       2  0 12:33 ?        00:00:00 [kworker/0:0-events]
2835    2430 18 12:33 pts/1    00:00:00 critical_application.py
2836    1875  0 12:33 pts/0    00:00:00 ps -ef
\end{verbbox}
\resizebox{8.6cm}{4.5cm}{\fbox{\theverbbox}}
\caption{(Step 1) Process list before Victim model was terminated. Victim's pid is observed to be 2835.}
\label{fig:a5}
\end{figure}

\begin{figure}[t!]
\begin{verbbox}[\scriptsize\slshape]
1874       1  0 Jul12 ?        00:00:00 /usr/sbin/dropbear -i -r /etc/dropbear/dropbear_rsa_host_key -B
1875    1874  0 Jul12 pts/0    00:00:00 -sh
2056       2  0 Jul12 ?        00:00:00 [kworker/3:2-cgroup_destroy]
2429       1  0 05:11 ?        00:00:00 /usr/sbin/dropbear -i -r /etc/dropbear/dropbear_rsa_host_key -B
2430    2429  0 05:11 pts/1    00:00:00 -sh
2456       2  0 05:14 ?        00:00:00 [kworker/2:0H]
2457       2  0 05:14 ?        00:00:00 [DPUCZDX8G_2]
2458       2  0 05:14 ?        00:00:00 [DPUCZDX8G_1]
2776       2  0 11:39 ?        00:00:00 [kworker/u8:1-events_unbound]
2802       2  0 11:55 ?        00:00:00 [kworker/3:1H]
2803       2  0 11:55 ?        00:00:00 [kworker/1:2H]
2811       2  0 12:04 ?        00:00:00 [kworker/0:2-events_power_efficient]
2819       2  0 12:16 ?        00:00:00 [kworker/u8:0-events_unbound]
2824       2  0 12:21 ?        00:00:00 [kworker/0:0-mm_percpu_wq]
2825     849  0 12:25 ?        00:00:00 systemd-userwork
2826     849  0 12:25 ?        00:00:00 systemd-userwork
2827     849  0 12:25 ?        00:00:00 systemd-userwork
2828       2  0 12:27 ?        00:00:00 [kworker/0:1-events]
2829       2  0 12:29 ?        00:00:00 [kworker/u8:2-events_unbound]
2830    1875  0 12:30 pts/0    00:00:00 ps -ef
\end{verbbox}
\resizebox{8.6cm}{4.5cm}{\fbox{\theverbbox}}
\caption{(Step 1) Process list after victim model is terminated.}
\label{fig:a6}
\end{figure}

\begin{figure}[t!]
\begin{verbbox}[\scriptsize\slshape]
1874       1  0 Jul12 ?        00:00:00 /usr/sbin/dropbear -i -r /etc/dropbear/dropbear_rsa_host_key -B
1875    1874  0 Jul12 pts/0    00:00:00 -sh
2056       2  0 Jul12 ?        00:00:00 [kworker/3:2-cgroup_destroy]
2429       1  0 05:11 ?        00:00:00 /usr/sbin/dropbear -i -r /etc/dropbear/dropbear_rsa_host_key -B
2430    2429  0 05:11 pts/1    00:00:00 -sh
2456       2  0 05:14 ?        00:00:00 [kworker/2:0H]
2457       2  0 05:14 ?        00:00:00 [DPUCZDX8G_2]
2458       2  0 05:14 ?        00:00:00 [DPUCZDX8G_1]
2776       2  0 11:39 ?        00:00:00 [kworker/u8:1-events_unbound]
2802       2  0 11:55 ?        00:00:00 [kworker/3:1H]
2803       2  0 11:55 ?        00:00:00 [kworker/1:2H]
2811       2  0 12:04 ?        00:00:00 [kworker/0:2-events_power_efficient]
2819       2  0 12:16 ?        00:00:00 [kworker/u8:0-events_unbound]
2824       2  0 12:21 ?        00:00:00 [kworker/0:0-mm_percpu_wq]
2825     849  0 12:25 ?        00:00:00 systemd-userwork
2826     849  0 12:25 ?        00:00:00 systemd-userwork
2827     849  0 12:25 ?        00:00:00 systemd-userwork
2828       2  0 12:27 ?        00:00:00 [kworker/0:1-events]
2829       2  0 12:29 ?        00:00:00 [kworker/u8:2-events_unbound]
2830    1875  0 12:30 pts/0    00:00:00 ps -ef
2840    2430 18 12:33 pts/1    00:00:00 python3 adversary.py
\end{verbbox}
\resizebox{8.6cm}{4.5cm}{\fbox{\theverbbox}}
\caption{(Step 2) Process list after adversary start execution. Adversary id is seen to be 2840.}
\label{fig:a7}
\end{figure}

\vspace{1ex}
\noindent\textbf{Step 3. Scraping virtual addresses:} In the second step of the process, the adversary requests initiation of memory protection by isolation. This changes the ID value in the XMPU configuration register from the victim's to the adversary's, while leaving the memory addresses in the XMPU configuration unchanged (retaining the victim's settings). 

It is important to note that when a process requests isolation via the embedded OS, it is assigned virtual memory addresses, which is mapped to the physical addresses pre-configured in the XMPU registers. Consequently, physical addresses remain invariant, even though the virtual addresses may change. Every process that requests isolation uses the same physical addresses for its execution in the FPGA's onboard memory unless the physical addresses in the XMPU registers are reconfigured differently by a secure master. Since these registers are not accessible from the user space, the adversary does not know the physical addresses at this point.

However, the adversary can simply requests isolation and learn the virtual addresses assigned to it by PetaLinux. In Step 4, we describe how these virtual addresses can then be converted to a physical addresses.  This conversion reveals the physical memory locations that the victim previously used, as they are the same physical addresses that were initially configured in the XMPU registers.

To obtain critical information about the virtual address space, adversary executes the command {\sf vim /proc/2840/maps}, which enables adversary to access the memory map of the process with the {\sf PID 2840}. Figure \ref{fig:a8} provides adversary with valuable insight into the heap's virtual address range, specifically ranging from {\sf 0xaaab0fcf1000} to {\sf 0xaaab11306000}. This range signifies the virtual address space used and allocated by the victim model for storing its data. 

\begin{figure}[t!]
\begin{verbbox}[\scriptsize\slshape]
aaab0fcf1000-aaab11306000 rw-p 00000000 00:00 0            [heap]
ffffa52a5000-ffffa6b2f000 rw-p 00000000 00:00 0
ffffa8035000-ffffa805b000 rw-s 00000000 00:05 733          /dev/dri/renderD128
\end{verbbox}
\resizebox{8.6cm}{0.9cm}{\fbox{\theverbbox}}
\caption{(Step 3) Virtual address of the adversary process ranges from 0xaaab0fcf1000 to 0xaaab11306000 in the heap.}
\label{fig:a8}
\end{figure}
\begin{figure}[t!]
\centering
\begin{verbbox}[\scriptsize\slshape]
aaab0fd8ef10: 0000 0000 0000 0000 0000 0000 0000 0000  ................
aaab0fd8ef20: ffff ffff ffff ffff ffff ffff ffff ffff  ................
aaab0fd8ef30: ffff ffff ffff ffff ffff ffff ffff ffff  ................
aaab0fd8ef40: ffff ffff ffff ffff ffff ffff ffff ffff  ................
aaab0fd8ef50: ffff ffff ffff ffff ffff ffff ffff ffff  ................
aaab0fd8ef60: ffff ffff ffff ffff ffff ffff ffff ffff  ................
aaab0fd8ef70: ffff ffff ffff ffff ffff ffff ffff ffff  ................
...
...
...
aaab0fd8f0d0: ffff ffff ffff ffff ffff ffff ffff ffff  ................
aaab0fd8f0e0: ffff ffff ffff ffff ffff ffff ffff ffff  ................
aaab0fd8f0f0: ffff ffff ffff ffff ffff ffff ffff ffff  ................
aaab0fd8f100: ffff ffff ffff ffff ffff ffff ffff ffff  ................
aaab0fd8f110: ffff ffff ffff ffff ffff ffff ffff fffe  ................
aaab0fd8f120: ffff fcff fffc ffff feff fffe fffe ffff  ................
aaab0fd8f130: feff fffd ffff feff ffff fff4 f7f5 f8fd  ................
\end{verbbox}
\resizebox{8.6cm}{4.5cm}{\fbox{\theverbbox}}
\caption{(Step 4) This figure illustrates a hexadecimal dump of the victim's data (ffff ffff) located in the virtual memory addresses ranging from 0xaaab0fd8ef20 to 0xaaab0fd8f130.}
\label{fig:a9}
\end{figure}

\vspace{1ex}
\noindent\textbf{Step 4. Mapping virtual addresses to physical addresses}: Adversary executes the command {\sf xxd} to perform a hexdump of the heap memory associated with the specified process ID (PID). The resulting output is saved to a file for further analysis. By using the {\sf grep} command and searching for the pixel values {\sf ffff ffff,} which represent the data written by victim application ({\sf critical\_application.py}). Figure \ref{fig:a9} shows the memory range encompassing the victims writes of 0xffffffff, spanning from memory addresses {\sf 0xaaab0fd8ef20} to {\sf 0xaaab0fd8f130}. To translate these virtual memory addresses to physical addresses, we refer to the information obtained from the {\sf page maps} file associated with the PID.  Thus, adversary obtains a range of {\sf 0x70c6df20} to {\sf 0x70c6e130} for physical addresses. 

\vspace{1ex}
\noindent\textbf{Step 5. Scraping memory residue:} To read data from physical addresses, adversary uses the command {\sf devmem \textlangle physical\_address\textrangle }. Thus, adversary retrieves data from the physical addresses obtained in Step 4. In order to demonstrate that adversary was able to retrieve the victim's data {\sf 0xffffffff} (corrupted image) from these physical locations, adversary ran the {\sf devmem} command for each individual physical memory location, ranging from {\sf 0x70c6df20} to {\sf 0x70c6e130}. The results of each {\sf devmem} command execution are illustrated in Figure \ref{fig:a10}. However, since the steps are automated, the {\sf devmem} command is executed for all the physical address locations specified in Step 4.
\begin{figure}[t!]
\begin{verbbox}[\scriptsize\slshape]
xilinx-zcu102-20222:~/READ# devmem 0x70c6df20        
0xFFFFFFFF
xilinx-zcu102-20222:~/READ# devmem 0x70c6df24       
0xFFFFFFFF
xilinx-zcu102-20222:~/READ# devmem 0x70c6df28              
0xFFFFFFFF
xilinx-zcu102-20222:~/READ# devmem 0x70c6df2c
0xFFFFFFFF
...
...
...
xilinx-zcu102-20222:~/READ# devmem 0x70c6e130
0xFEFFFFFD
xilinx-zcu102-20222:~/READ# devmem 0x70c6e124
0xFFFFFEFF
xilinx-zcu102-20222:~/READ# devmem 0x70c6e128
0xFFFFFFF4
xilinx-zcu102-20222:~/READ# devmem 0x70c6df88
0xF7F5F8FD
\end{verbbox}
\resizebox{8.6cm}{4.5cm}{\fbox{\theverbbox}}
\caption{(Step 5) The figure shows the data read is 0xFFFFFF which identifies that the read out is of victim's.}
\label{fig:a10}
\end{figure}

\begin{figure}[t!]
\begin{verbbox}[\scriptsize\slshape]
aaaaf1676000: 0000 0000 0000 0000 9102 0000 0000 0000  ................
aaaaf1676010: 0700 0200 0200 0200 0100 0100 0700 0700  ................
aaaaf1676020: 0600 0400 0600 0600 0300 0400 0000 0700  ................
aaaaf1676030: 0000 0400 0200 0700 0600 0100 0000 0000  ................
aaaaf1676040: 0000 0100 0200 0200 0200 0300 0100 0300  ................
aaaaf1676050: 0500 0000 0300 0100 0500 0100 0200 0000  ................
aaaaf1676060: 0000 0100 0200 0000 0000 0300 0000 0000  ................
aaaaf1676070: 0000 0000 0000 0000 0000 0100 0000 0000  ................
aaaaf1676080: 0000 0000 0000 0000 0000 0000 0000 0000  ................
aaaaf1676090: 8007 71f1 aaaa 0000 7012 71f1 aaaa 0000  ..q.....p.q.....
....
....
....
aaaaf1713e60: 0000 0000 0000 0000 0000 0000 0000 0000  ................
aaaaf1713e70: 5555 5555 5555 5555 5555 5555 5555 5555  UUUUUUUUUUUUUUUU
aaaaf1713e80: 5555 5555 5555 5555 5555 5555 5555 5555  UUUUUUUUUUUUUUUU
aaaaf1713e90: 5555 5555 5555 5555 5555 5555 5555 5555  UUUUUUUUUUUUUUUU
aaaaf1713ea0: 5555 5555 5555 5555 5555 5555 5555 5555  UUUUUUUUUUUUUUUU
aaaaf1713eb0: 5555 5555 5555 5555 5555 5555 5555 5555  UUUUUUUUUUUUUUUU
aaaaf1713ec0: 5555 5555 5555 5555 5555 5555 5555 5555  UUUUUUUUUUUUUUUU
aaaaf1713ed0: 5555 5555 5555 5555 5555 5555 5555 5555  UUUUUUUUUUUUUUUU
aaaaf1713ee0: 5555 5555 5555 5555 5555 5555 5555 5555  UUUUUUUUUUUUUUUU
aaaaf1713ef0: 5555 5555 5555 5555 5555 5555 5555 5555  UUUUUUUUUUUUUUUU
aaaaf1713f00: 5555 5555 5555 5555 5555 5555 5555 5555  UUUUUUUUUUUUUUUU
aaaaf1713f10: 5555 5555 5555 5555 5555 5555 5555 5555  UUUUUUUUUUUUUUUU
aaaaf1713f20: 5555 5555 5555 5555 5555 5555 5555 5555  UUUUUUUUUUUUUUUU
aaaaf1713f30: 5555 5555 5555 5555 5555 5555 5555 5555  UUUUUUUUUUUUUUUU
aaaaf1713f40: 5555 5555 5555 5555 5555 5555 5555 5555  UUUUUUUUUUUUUUUU
aaaaf1713f50: 5555 5555 5555 5555 5555 5555 5555 5555  UUUUUUUUUUUUUUUU
aaaaf1713f60: 5555 5555 5555 5555 5555 5555 5555 5555  UUUUUUUUUUUUUUUU
aaaaf1713f70: 5555 5555 5555 5555 5555 5555 5555 5555  UUUUUUUUUUUUUUUU
aaaaf1713f80: 5555 5555 5555 5555 5555 5555 5555 5555  UUUUUUUUUUUUUUUU
aaaaf1713f90: 5555 5555 5555 5555 5555 5555 5555 5555  UUUUUUUUUUUUUUUU
aaaaf1713fa0: 5555 5555 5555 5555 5555 5555 5555 5555  UUUUUUUUUUUUUUUU
aaaaf1713fb0: 5555 5555 5555 5555 5555 5555 5555 5555  UUUUUUUUUUUUUUUU
aaaaf1713fc0: 5555 5555 5555 5555 5555 5555 5555 5555  UUUUUUUUUUUUUUUU
aaaaf1713fd0: 5555 5555 5555 5555 5555 5555 5555 5555  UUUUUUUUUUUUUUUU
aaaaf1713fe0: 5555 5555 5555 5555 5555 5555 5555 5555  UUUUUUUUUUUUUUUU
aaaaf1713ff0: 5555 5555 5555 5555 5555 5555 5555 5555  UUUUUUUUUUUUUUUU
aaaaf1714000: 5555 5555 5555 5555 5555 5555 5555 5555  UUUUUUUUUUUUUUUU
aaaaf1714010: 5555 5555 5555 5555 5555 5555 5555 5555  UUUUUUUUUUUUUUUU
aaaaf1714020: 5555 5555 5555 5555 5555 5555 5555 5555  UUUUUUUUUUUUUUUU
aaaaf1714030: 5555 5555 5555 5555 5555 5555 5555 5555  UUUUUUUUUUUUUUUU
aaaaf1714040: 5555 5555 5555 5555 5555 5555 5555 5555  UUUUUUUUUUUUUUUU
aaaaf1714050: 5555 5555 5555 5555 5555 5555 5555 5654  UUUUUUUUUUUUUUVT
aaaaf1714060: 5556 5255 5652 5755 5457 5554 5754 5657  UVRUVRWUTWUTWTVW

\end{verbbox}
\resizebox{8.6cm}{7.5cm}{\fbox{\theverbbox}}
\caption{(Step 4) Experiment with victim writing 0x55555555 into memory.}
\label{fig:a11}
\end{figure}

\vspace{1ex}
\textbf{Step 6. Analysis of data scraped from victim:}  In this example, we as adversary have knowledge of what the victim was writing, allowing us to identify it in the hexdump during Step 4. From there, we converted those virtual addresses to physical addresses and proceeded to Step 5 to scrape memory residue from these addresses. However, in a real-world scenario, the adversary may not have prior knowledge of the type of application the victim is running or the domain it is associated with. Therefore, the adversary needs to identify the domains for which this board is being used as hardware accelerators and work on obtaining reference files from those domains for effective comparison, as explained in Section 3 in Step 6 - \textit{Profiling} and \textit{Reconstruction}.

\vspace{1ex}
Following the steps described in Section 3, we have shown that XMPU does not sanitize memory residue left by a terminated victim in both scenarios described under \textit{Adversary Process} in Section 4. If the adversary has prior knowledge of the victim applications, it can mount a potent attack on the memory residue to reconstruct the victim usage scenario.

Our experiments did not stop at the victim writing {\sf 0xffffffff}; we also tested various other identifiers, such as {\sf 0x55555555}, as shown in Figure \ref{fig:a11}, to validate the robustness of our approach. These additional tests reaffirmed the effectiveness of our approach in exposing the XMPU security vulnerabilities.
\section{conclusion}
The Xilinx Memory Protection Unit promises protection for active processes by isolating them from each other. However, as this paper shows, XMPU does not clear the memory for terminated processes, leaving it exposed to attacker processes even when the victim process ran under XMPU protection. It is worth noting that memory reconfiguration solely takes place during the First Stage Boot Loader (FSBL) phase, immediately following a power-on reset. Erasing memory upon process termination can lead to a performance overhead and increased reconfiguration time, which may explain Xilinx's decision not to do so. However, as this paper has shown, failing to clear memory leaves private data belonging to the terminated process exposed to other processes. This is a serious security flaw in the XMPU architecture.


\bibliographystyle{unsrt}
\bibliography{references}

\end{document}